\documentclass[10pt]{article}
\usepackage{graphicx}

\def\Title#1{\begin{center} {\Large #1 } \end{center}}
\def\Author#1{\begin{center}{ \sc #1} \end{center}}
\def\Address#1{\begin{center}{ \it #1} \end{center}}

\newcommand\pubblock{\rightline{\begin{tabular}{l} Proceedings of the Fifth Annual LHCP\\ \pubnumber\\
         \pubdate  \end{tabular}}}

\newenvironment{Abstract}{\begin{quotation} \begin{center} 
             \large ABSTRACT \end{center}\bigskip 
      \begin{center}\begin{large}}{\end{large}\end{center} \end{quotation}}

\newenvironment{Presented}{\begin{quotation} \begin{center} 
             PRESENTED AT\end{center}\bigskip 
      \begin{center}\begin{large}}{\end{large}\end{center} \end{quotation}}

\def\Acknowledgements{\bigskip  \bigskip \begin{center} \begin{large}
             \bf ACKNOWLEDGEMENTS \end{large}\end{center}}

\def\beq{\begin{equation}}
\def\eeq#1{\label{#1}\end{equation}}
\def\eeqn{\end{equation}}

\def\beqa{\begin{eqnarray}}
\def\eeqa#1{\label{#1}\end{eqnarray}}
\def\eeqan{\end{eqnarray}}

\let\bar=\overbar

\def\Dslash{\not{\hbox{\kern-4pt $D$}}}
\def\dslash{\not{\hbox{\kern-2pt $\del$}}}

\def\msb{{\bar{\ssstyle M \kern -1pt S}}}

\usepackage{color}

\newcommand\pubnumber{LHCb-PROC-2017-033}

\newcommand\pubdate{September 7, 2017}

\def\affiliation{
On behalf of the LHCb Experiment, \\
Physikalisches Institut \\
University of Heidelberg, Germany}

\begin{document}

\large
\begin{titlepage}
\pubblock

\vfill
\Title{New results on collectivity with LHCb}
\vfill

\Author{ Renata Kope\u{c}n\'{a}  }
\Address{\affiliation}
\vfill
\begin{Abstract}

Two-particle angular correlations are studied in $pp$ collisions at $\sqrt{s}=$\,13\,TeV, collected with the LHCb detector at the LHC. 
The LHCb detector provides measurement in the very forward region, $2 < \eta < 5$. This region is 
complementary to other experiments and allows to explore low Bjorken-$x$ region. The correlations
are studied  as a function of difference in pseudorapidity 
$\left(\Delta\eta\right)$ and azimuthal angle $\left(\Delta\phi\right)$ in several $p_T$ and activity classes. 
Proton-proton collisions are studied using two datasets corresponding to minimum-bias sample and a sample
obtained via a  dedicated trigger to study the highest-activity events.

\end{Abstract}
\vfill

\begin{Presented}
The Fifth Annual Conference\\
 on Large Hadron Collider Physics \\
Shanghai Jiao Tong University, Shanghai, China\\ 
May 15-20, 2017
\end{Presented}
\vfill
\end{titlepage}
\def\thefootnote{\fnsymbol{footnote}}
\setcounter{footnote}{0}
%

\normalsize 


\section{Introduction}

Two-particle correlations are one of the tools to investigate heavy-ion collisions and its collective effects. Azimuthal anisotropies of particles produced in such collisions are one of the signatures of Quark Gluon Plasma (QGP). QGP is very hot and dense state of matter, where quarks and gluons are deconfined. 

The experiments ALICE~\cite{Abelev:2013kop}, ATLAS~\cite{Aad:2016kop},  CMS~\cite{Khachatryan:2016kop}, LHCb~\cite{Aaij:2015kop} and STAR~\cite{Abelev:2016kop} presented their results suggesting the two-particle correlation function of low-multiplicity $PbPb$ events evinces same features as $pPb$ and even $pp$ events. LHCb can provide unique results in the $\eta$ range $ 2 < \eta < 5$, complimentary to other LHC experiments. The forward pseudorapidity region is in the center of interest due to its ability to explore small Bjorken-$x$ region and hence probe theories such as Color-Glass Condensate (CGC). 

The study of the correlation function $C\left(\Delta\eta,\Delta\phi\right)$ is conventionally done in two dimensions as a function of separation in pseudorapidity, $\Delta\eta$, and azimuthal angle, $\Delta\phi$, for pairs of charged prompt particles.  This function includes many features in different regions. The most striking feature is near-side peak, originating from jets. This peak is more narrow with increasing $p_T$. 

The away side ridge is the region of the correlation function $\Delta\phi\,\approx\,\pi$. Its presence is caused by back-to-back jets. This ridge is apparent in the whole $p_T$ range. The away-side ridge is expected to be more pronounced in $pp$ a $pPb$ collisions compared to heavy ion collisions due to the jet quenching effect~\cite{Adams:2003kop}. 
	
Due to the isotropic resonance decays contributing in the whole $\Delta\phi$ region, the Gaussian ridge, stretching
along the whole $\Delta\phi$ at $\Delta\eta\,\approx\,0$ , is present.

\subsection{Near side ridge}

The so-called near side ridge is currently in the center of interest of many experimentalists and theorists.
The near side ridge ($\Delta\phi\,\approx\,0$) was observed first in $Au + Au$ collisions~\cite{Adams:2003kop}. 
In such heavy-ion collisions, the
near side ridge is very pronounced and assumed to be a result of strongly-interacting QGP. However, the
near side ridge was observed also later in  $p + Pb$ collisions\,\cite{Khachatryan:2016kop}.

The discussion of its origin is still ongoing. Many models tried to describe this
phenomenon, including CGC, collectivity or jet-medium interactions \cite{Dusling1:2013kop}\,-\,\cite{Shuryak:2013kop}. The unique
acceptance of LHCb provides new input to these theories and deepen the understanding of underlying mechanisms in high-energy collisions.

\section{Experimental setup}

The two-particle angular correlations are studied in $pp$ collisions at $\sqrt s$ = 13 TeV, collected with the LHCb
detector at the LHC. The LHCb detector~\cite{Alves:2008kop}\,,\cite{Aaij:2014kop} 
is a single-arm forward spectrometer covering the pseudorapidity
range $2 < \eta < 5$, designed to study particles containing b or c quarks. The detector includes
a high-precision tracking system consisting of a silicon-strip vertex detector surrounding
the $pp$ interaction region~\cite{Aaij:2014VELOkop}, a large-area silicon-strip detector located upstream of a dipole magnet with a bending power of about 4 Tm, and three stations of silicon-strip
detectors and straw drift tubes~\cite{Arink:2013kop}  placed downstream of the magnet. The tracking system provides a measurement of momentum, p, with a relative uncertainty that varies from 0.5\%
at low values to 1.0\% at 200 GeV/c. The minimum distance of a track to a primary vertex, the impact parameter (IP), is measured with a resolution of
$\left(15 + 29  p_T \right) $ $\mu$m, where $p_T$ is the component of the momentum transverse to the beam, in GeV/c. Different types of
charged hadrons are distinguished using information from two ring-imaging Cherenkov detectors \cite{Adinolfi:2013kop}. Photons, electrons and hadrons are identified by a calorimeter system consisting of scintillating-pad and preshower detectors, an electromagnetic calorimeter (ECAL) and a hadronic calorimeter (HCAL). Muons are identified by a system composed of alternating layers of iron and multiwire proportional chambers \cite{Alves:2012kop}.

Two trigger settings were used: a minimum bias trigger, selecting the first bunch crossing in a train
to avoid spillover and parasitic collisions and a dedicated high-activity trigger selecting high activity events
corresponding to ∼ 1\% of activity. The latter trigger setting selected only events with a reconstructed vertex requiring at least 4 tracks.

For efficiency corrections, a Monte Carlo sample of $pp$ collisions at $\sqrt s = 13$\,TeV was used. The $pp$ collisions were generated using \textsc{Pythia 8}  with specific LHCb configuration \cite{Belyaev:2011kop}. The interaction of the
generated particles with the detector, and its response are implemented using the \textsc{Geant4} toolkit \cite{Geant4:2003kop} as described in \cite{Clemencic:2011kop}.

\section{Analysis method}

Each event is required to have exactly one reconstructed primary vertex. The reconstructed vertex is required
to be near the mean interaction point in order to remove beam-related background interaction. To further
suppress beam-gas interactions, beam splashes and events where protons migrated from a nominal to empty
bunch, events with a small ratio of hits in the ECAL and VELO are rejected.

The event-activity was determined using hit-multiplicities in the VELO $ N_{hit}^{VELO}$. This variable was chosen because
the VELO covers pseudorapidity, $-4.0 < \eta < -1.5$; $1.5 < \eta < 5.0$ \cite{Aaij:2014VELOkop} and other subdetectors cover only
sub-ranges of $\eta$. Moreover, hit-multiplicity is not biased by reconstruction algorithms.

Only charged tracks traversing the full LHCb tracking system were used. Candidates were required to
have $p > 2$\,GeV, $p_T > 150$\,MeV and $1.9 < \eta < 4.9$. To suppress background induced by decaying particles,
tracks were required to pass quality cuts and to have small distance to the reconstructed primary vertex.
he study was done in four multiplicty clases and four momentum classes.

Due to the limitations in acceptance and reconstruction of the tracks, corrections are applied per track.
Moreover, corrections for non-prompt particles and fake-track contaminations are applied. The correction is done in 4D 
in $\left( \eta, p_T, \phi, N_{hit}^{VELO}  \right)$.

\section{Results}

 The signal function is a 2D function of the pair-wise difference in $\eta$ and $\phi$ for all particles in one event. The signal
distribution is defined as:
  \begin{equation}\label{eq:sig}
        S\left(\Delta\eta,\Delta\phi\right) = \frac{1}{N_{trig}} \frac{d^2N_{same}}{d\Delta\eta d\Delta\phi}\,.         
  \end{equation}

The number of trigger particles, $N_{trig}$, denotes the number of all particles in the given $p_T$ and activity bin.
This rather confusing term originates from the the name of a particle, \emph{trigger particle}, that is correlated with
\emph{associated particles}. $N_{same}$ denotes pairs of tracks correlated within the same event.

The signal distribution is distorted by the limited acceptance and reconstruction artifacts. Therefore, the signal 
distribution is normalized by the \emph{background distribution}. The background distribution $B\left(\Delta\eta,\Delta\phi\right)$
is defined as:
  \begin{equation}\label{eq:background}
    B\left(\Delta\eta,\Delta\phi\right) = \frac{d^2N_{mix}}{d\Delta\eta d\Delta\phi}\,.
  \end{equation}

This time, $N_{mix}$ stands for the number of particle pairs obtained by combining a trigger particle from one
event with all the tracks from similar events. The mixed events were chosen to be within 1\% of the centrality
class for the minimum-bias data sample and 0.1\% for the high-activity sample. In order to normalize
this distribution, $B\left(\Delta\eta,\Delta\phi\right)$ is divided by the zeroth bin $B\left(0,0\right)$. Given this normalization, background
distribution can be understood as a pair-acceptance efficiency.

The correlation function itself is obtained by:
  \begin{equation}\label{eq:result}
    \frac{1}{N_{trig}} \frac{d^2N_{pair}}{d\Delta\eta d\Delta\phi} = B\left(0,0\right) \times \frac{ S\left(\Delta\eta,\Delta\phi\right)}{ B\left(\Delta\eta,\Delta\phi\right)}\,.
  \end{equation}
The included background makes this analysis rather robust. The tracking and acceptance effects are minimized using this approach.

\begin{figure}[htb]
\centering
\includegraphics[height=2.75in]{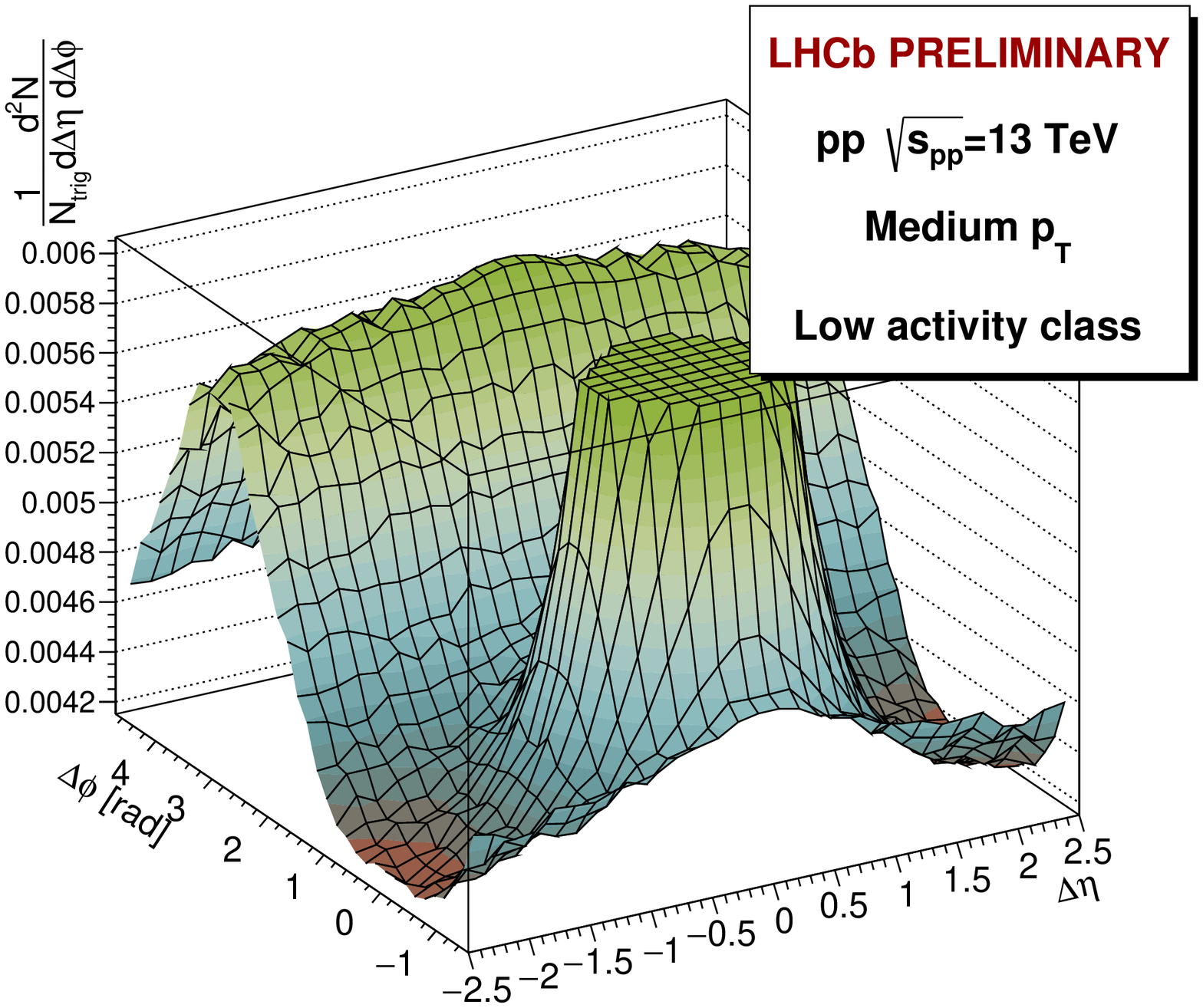}
\includegraphics[height=2.75in]{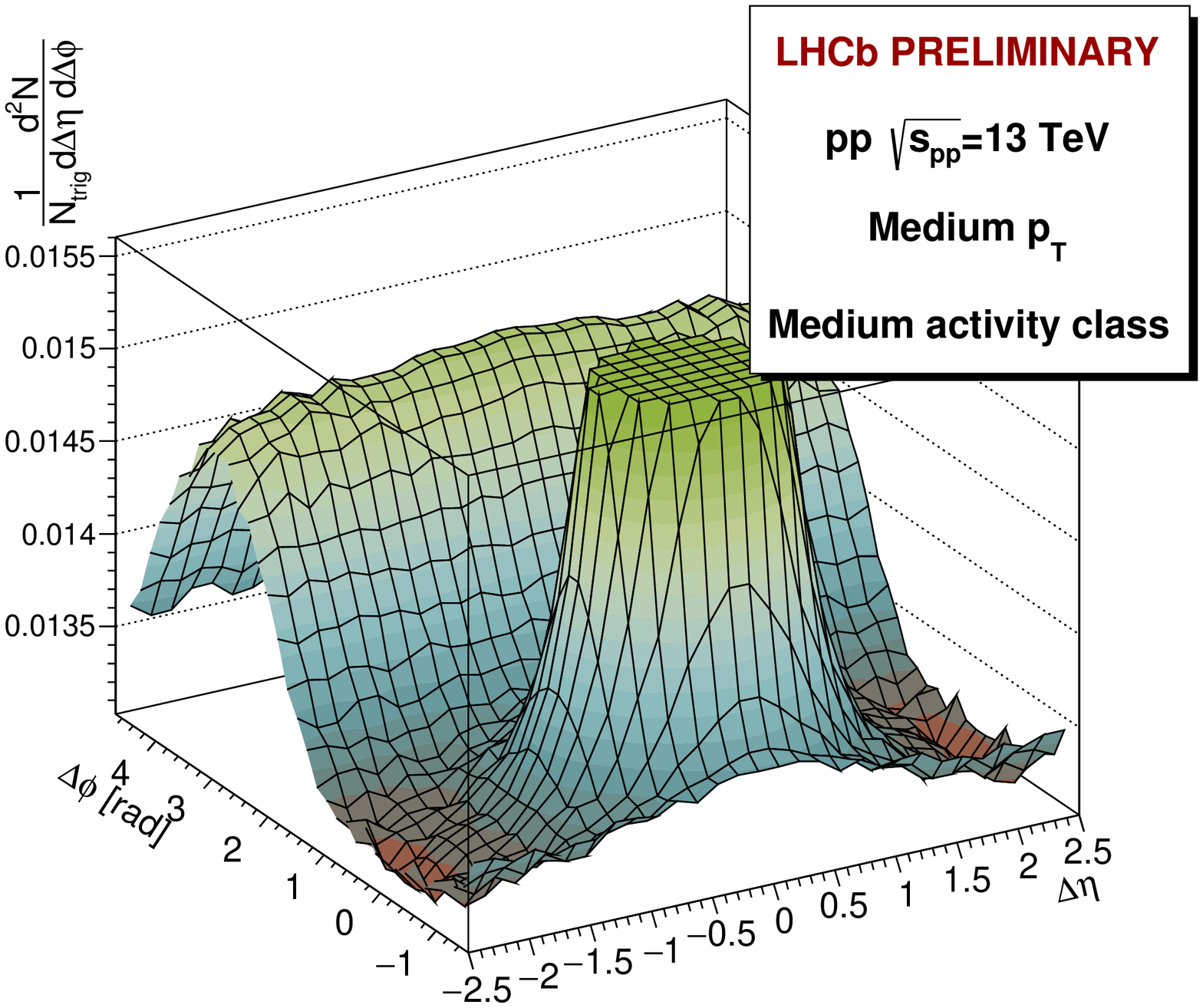}
\caption{Two particle correlation function for the low and medium activity class and medium $p_T$ class.
The top of the near side peak is cut off for viewers comfort.}
\label{fig:kop_fig1}
\end{figure}

\begin{figure}[htb]
\centering
\includegraphics[height=2.75in]{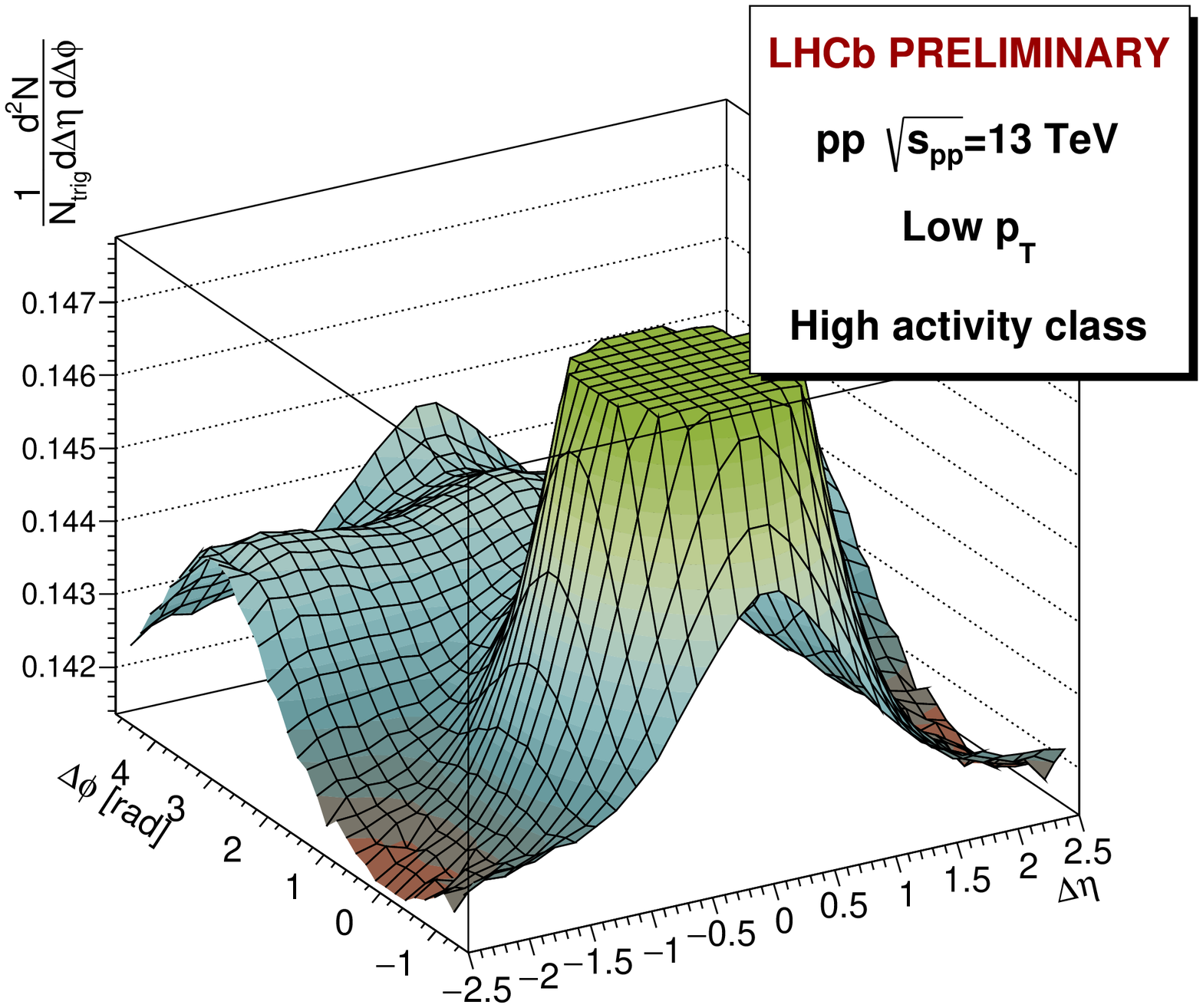}
\includegraphics[height=2.75in]{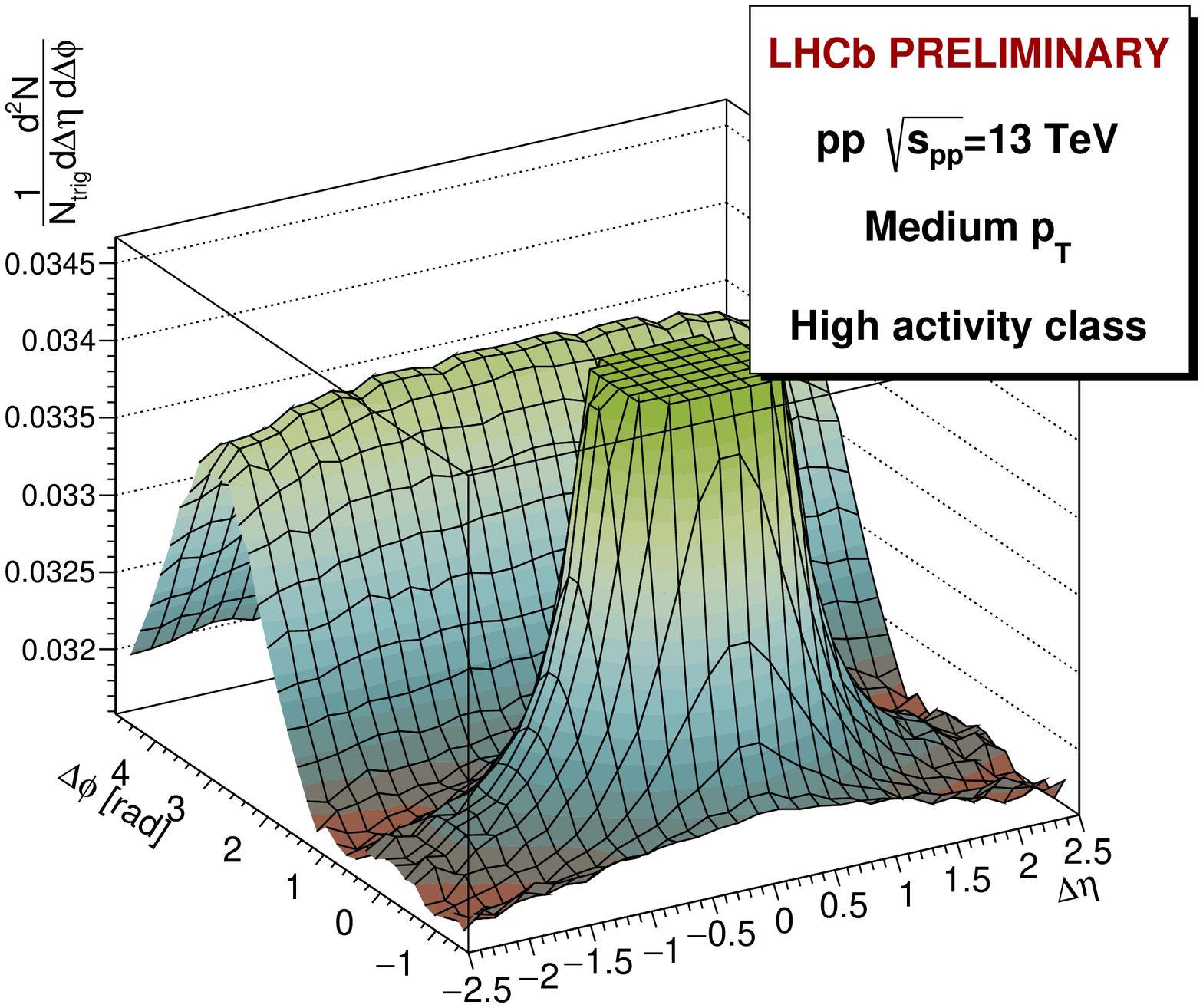}
\caption{Two particle correlation function for the high activity and low and medium  $p_T$ class.
The top of the near side peak is cut off for viewers comfort.}
\label{fig:kop_fig2}
\end{figure}

The correlation functions for several $p_T$ and activity classes are in the Figure \ref{fig:kop_fig1} and \ref{fig:kop_fig2}. 
For low $p_T$, the Gaussian ridge is clearly pronounced due to isotropic resonance decay. The near-side peak is very broad
due to the fact that the decay products of low $p_T$ particles are less collimated. The away-side ridge is the most pronounced
due to low $p_T$ resonance decays. At high $p_T$, the near side peak is more narrow. The away side ridge is less pronounced and the Gaussian ridge disappears. 

At the medium activity, no hints of a near side ridge appear. However, at medium $p_T$ and high activity a hint of near side ridge appears. 
This provides motivation for further studies.

\section{Conclusions}

For the first time, two-particle correlations 
between charged prompt particles produced in $pp$ collisions 
at $\sqrt{s} = 13$\,TeV were measured using the LHCb detector. 
The angular correlations were studied in the very forward region $2 < \eta < 5$,
complimenting previous LHC studies, and over the full azimuthal angle. 
The correlation function was studied in different $p_T$
and activity intervals.

\Acknowledgements
I would like to thank S. Stahl, F. Bossu, M. Winn, W. Barter and X. Cid Vidal for stimulating discussions.
I  express  my  gratitude to our colleagues in the CERN accelerator departments for the excellent 
performance of the LHC. I thank the technical and administrative staff at the LHCb institutes. 
I wish to thank the Heidelberg Graduate School of Fundamental Physics (HGSFP) for financial support.

\end{document}